\documentclass[aps,showpacs,twocolumn]{revtex4}
\usepackage[dvips]{graphicx}
\usepackage{amssymb}
\usepackage{bm}

\newcommand{\be}{\begin{equation}}
\newcommand{\ee}{\end{equation}}
\newcommand{\bear}{\begin{eqnarray}}
\newcommand{\eear}{\end{eqnarray}}
\newcommand{\ba}{\begin{array}}
\newcommand{\ea}{\end{array}}
\newcommand{\nn}{\nonumber}

\begin{document}
\setcounter{page}{1}
\title[]{Lower Dimensional Branes in Boundary Conformal Field Theory}
\author{Akira \surname{Ishida}} \email{ishida@skku.edu}
\author{Yoonbai \surname{Kim}} \email{yoonbai@skku.edu}
\affiliation{Department of Physics
and Institute of Basic Science, Sungkyunkwan University,
Suwon 440-746}
\author{Chanju \surname{Kim}} \email{cjkim@ewha.ac.kr}
\affiliation{Department of Physics, Ewha Womans University, Seoul 120-750}
\author{O-Kab \surname{Kwon}} \email{okabkwon@maths.tcd.ie}
\affiliation{Center for Quantum Spacetime, Sogang University, Seoul 121-742}
\affiliation{School of Mathematics, Trinity College, Dublin, Ireland}


\begin{abstract}
In the presence of constant background electromagnetic fields, we discuss
three types of exactly marginal boundary tachyon operators
for static kinks in boundary conformal field theory.
Functional forms of three operators are hyperbolic sine, hyperbolic
cosine, and exponential types, and they describe codimension-one solitons
when the transverse electric field has overcritical value.
The energy-momentum tensor and the source for antisymmetric tensor
field are computed in the path integral approach
for the exponential-type tachyon vertex operator.
\end{abstract}



\maketitle

\section{Introduction}\label{section1}

Study of the open string tachyon has led to some understanding
of the nonperturbative aspects of string
theory~\cite{Sen:2004nf}. Among many results related to
the open string tachyon,
rolling tachyons and tachyon solitons have been important issues.

When an unstable D-brane decays homogeneously, the real-time decay
process of the D-brane can be described by a marginally-deformed
boundary conformal field theory (BCFT)~\cite{Sen:2002nu,Sen:2002in}.
A. Sen found one parameter
family of time-dependent solutions, referred to as rolling tachyon,
 is identified with the marginal deformation parameter.
This has been studied in Refs.~\cite{Sen:2002nu,Sen:2002in,Larsen:2002wc}
for the pure tachyon case and in Refs.~\cite{Mukhopadhyay:2002en,Rey:2003xs}
in the presence of background electromagnetic fields.
These time evolution properties of the unstable D-brane are also described
in the other effective field theories, {\it e.g.,} boundary string field
theory (BSFT)~\cite{Sugimoto:2002fp,Minahan:2002if,Ishida:2006mi},
Dirac-Born-Infeld (DBI)-type effective field theory
(EFT)~\cite{Mukhopadhyay:2002en,Sen:2002an, Kim:2003he},
and noncommutative effective field theory
(NCFT)~\cite{Mukhopadhyay:2002en,Kim:2005pz}.

Simultaneously, much work has been done on tachyon solitons,
for instance,
tachyon kinks~\cite{Sen:2003tm,Lambert:2003zr,Brax:2003rs,Kim:2003in,
Sen:2003bc,Banerjee:2004cw,Kim:2006mg}, tube~\cite{Kim:2003uc},
and vortices~\cite{Sen:2003tm,Sen:2003bc,Kim:2005tw}.
For the case of the pure tachyon, only the array of kink-antikink is
found~\cite{Lambert:2003zr,Kim:2003in,Brax:2003rs}.
In the presence of constant electromagnetic fields,
$F_{\mu\nu}$, rich spectra of the tachyon kinks are found
and those configurations are classified according to the sign of
$C^{11}$, the (11)-component of the cofactor for matrix
$(\eta +F)_{\mu\nu}$, where $x^1$ is the transverse direction to the
kink~\cite{Kim:2003in}.
When $C^{11}$ is negative, there are two species of kinks which are
the array of kink-antikink and the single topological BPS kink. These two
kinks are also identified in
BCFT, NCFT, and BSFT~\cite{Kim:2006mg}.
(In addition to the above two, there is another solution
in DBI-type EFT~\cite{Kim:2003in} and NCFT~\cite{Banerjee:2004cw}.)
On the other hand, for $p\ge 2$, $C^{11}$ can be positive when there are
nonzero components of electric field in both
longitudinal and transverse directions as well as nonzero magnetic field.
In this case, there are three more species of
kinks named bounce, half-kink, and topological non-BPS kink
in BSFT~\cite{Kim:2006mg}, DBI-type EFT~\cite{Kim:2003in},
and NCFT~\cite{Banerjee:2004cw}.
These solutions have not been investigated in BCFT.

The purpose of this paper is to study the latter three types of solutions
in the context of BCFT. We construct exactly marginal boundary tachyon
operators in the presence of constant electromagnetic fields.
Functional forms of these operators are hyperbolic cosine,
hyperbolic sine, and exponential types.
We compute the spacetime energy-momentum tensor and the source
for the antisymmetric tensor for the exponential-type operator
by using the path integral approach~\cite{Larsen:2002wc}
in bosonic string theory.

In II, we identify exactly marginal boundary tachyon operators
in the presence of constant background electromagnetic fields.
In III, we obtain the spacetime energy-momentum tensor for the
exponential-type operator by using the path integral approach.
We conclude in IV.

\section{Tachyon Vertices for Static Kinks}\label{section2}

Let us consider an unstable D$p$-brane of tension ${\cal T}_p\, (p\le 25)$
with fundamental strings in bosonic string theory.
Dynamics of the unstable D$p$-brane is described in BCFT by introducing
interaction terms of the tachyon $T(X)$ and the gauge field $A_\mu(X)$
on the boundary of the worldsheet.
Specifically, in the flat spacetime, the worldsheet action is given by
\begin{eqnarray}\label{SBCFT}
S_{\rm BCFT} = S_0+ S_A + S_T
\end{eqnarray}
with
\bear
S_0 &=& \frac1{2\pi}\int_{\Sigma}d^2w\,
\eta_{\mu\nu} \partial X^\mu\bar\partial X^\nu,
\nn \\
S_A&=&  \frac{i}{2\pi} \int_{\partial \Sigma}d\tau A_\mu (X)
\partial_\tau X^\mu,
\label{SA} \\
S_T &=&\int_{\partial \Sigma} d\tau\, T(X), \label{ST}
\eear
where $\Sigma$ denotes a worldsheet and $\tau$ is the coordinate along the
worldsheet boundary $\partial \Sigma$. We also have the deformed
boundary condition for $X^\mu (w,\bar w)$ in the presence of the
background electromagnetic field,
\be
(\partial - \bar\partial)X^\mu\mid_{w=\bar w}
 + F^\mu_{~~\lambda} (\partial + \bar\partial) X^\lambda
 \mid_{w=\bar w}\,=\,0,
\label{bnd1}
\ee
where
$w=\tau + i\sigma$ ($-\infty < \tau <\infty, \,\, 0\le \sigma
<\infty$) is the complex coordinate on the upper half plane (UHP).
From the action (\ref{SA}) we read the worldsheet energy-momentum tensor:
\be\label{Tww}
T_{ww} = -\eta_{\mu\nu} : \partial X^\mu (w)
\partial X^\nu (w) :\,.
\ee

The main purpose of this section is to find conformally-invariant
tachyon vertices, $S_{\rm T}$, in the
presence of a constant electromagnetic field $F_{\mu\nu}$,
which are appropriate for tachyon kinks.
For convenience, we take symmetric gauge, $A_\mu = -\frac12 F_{\mu\nu}X^\nu$.
The general exponential-type tachyon vertex operator
in open string theory is written as
\bear\label{tvo1}
T(X)
&=& \sum_i\left(\lambda^{i}_{+}e^{ik^{i}_{+}\cdot X}
+\lambda^{i}_{-}e^{-ik^{i}_{-} \cdot X} \right),
\eear
where $\lambda^i_{\pm}$ and $k^i_{\pm}$ should be chosen to
make $T(X)$ real.
Under the deformed boundary condition in Eq.~(\ref{bnd1}),
the correlation function on the upper half plane is obtained through
operator product expansion~\cite{Abouelsaood:1986gd}:
\bear\label{XX6}
&&\langle X^\mu(w) X^\nu(w')\rangle_{{\rm UHP}}
\nn \\
&&\hskip 1.0cm
=-\eta^{\mu\nu}\ln \mid w-w'\mid +\eta^{\mu\nu}\ln \mid w-\bar w'\mid
\nn \\
&&\hskip 1.3cm -G^{\mu\nu}\ln \mid w-\bar w'\mid^2
- \theta^{\mu\nu}\ln \left(\frac{w-\bar w'}{\bar w - w'}\right)
\eear
with open string metric $G^{\mu\nu}$ and noncommutative parameter
$\theta^{\mu\nu}$ defined by
\be
G^{\mu\nu} = \left(\frac{1}{\eta + F}\right)^{\mu\nu}_{{\rm S}},
\qquad
\theta^{\mu\nu} = \left(\frac{1}{\eta + F}\right)^{\mu\nu}_{{\rm A}},
\ee
where `S'(`A') denotes the symmetric (anti-symmetric)
part of the matrix component.
Then Eq.~(\ref{XX6}) on the boundary of coordinate $\tau'$ provides
\be \label{XX9}
X(w)X(\tau') \sim -G^{\mu\nu} \ln |w-\tau'|^2 -\theta^{\mu\nu}
\ln \left(\frac{w-\tau'}{\bar w -\tau'}\right).
\ee

The conformal weight of the tachyon vertex operator at the boundary
may be identified by considering the operator product expansion with
the worldsheet energy momentum-tensor. By using Eq.~(\ref{XX9}) and the
deformed boundary condition (\ref{bnd1}),
\bear\label{OPE2}
T_{ww}(w):e^{ik\cdot X(\tau')}:
&\sim&\frac{G^{\mu\nu}k_\mu k_\nu}{(w-\tau')^2}:e^{ik\cdot X(\tau')}:
\nn \\
&&+ \frac{1}{w-\tau'}\,\partial_{\tau'} :e^{ik\cdot X(\tau')}:.
\eear
Then the boundary operator has conformal weight $h = G^{\mu\nu}k_\mu k_\nu$
and becomes marginal when
\be\label{const1}
G^{\mu\nu}k_\mu k_\nu=1.
\ee

For the rest of the paper we will consider only the operators which
depend on a single spatial coordinate, say $X^1$, in bosonic string theory.
In this case, Eq.~(\ref{const1}) reduces to
\be\label{const4}
k_1^2 = \frac1{G^{11}} = \frac{Y}{C^{11}},
\ee
where $Y\equiv \det (\eta + F)$ and $C^{\mu\nu}$ is the cofactor of
the matrix $(\eta + F)_{\mu\nu}$.
For physical electromagnetic fields, $Y$ is nonpositive definite,
but $C^{11}$ can have both negative and positive values for $p\ge 2$.
Thus depending on the signature of $C^{11}$, $k_1$ can be real or imaginary.
This is in contrast with the case of rolling tachyons, in which
$k_0^2 = Y/C^{00}$ is always negative, since $Y$ is negative
and $C^{00}$ is kept positive.

When $C^{11}$ is negative ($k_1$ is real),
the exactly marginal tachyon vertex operator has the form,
up to a translation in $X^1$,
\be\label{Tkink1}
T(X) = \lambda \cos \left(\sqrt{\frac{-Y}{-C^{11}}} \, X^1\right),
\ee
where we set $\lambda_{+} = \lambda_{-} = \frac12\lambda$ and
$k_{+}^1 = k_{-}^1= k_1$ in Eq.~(\ref{tvo1}).
The resulting configuration for the pure tachyon case ($Y=C^{11}=-1$)
is interpreted as
an array of D($p-1$)$\bar {\rm D}(p-1)$~\cite{Sen:1999mh,Sen:1998tt}.
Eq.~(\ref{Tkink1}) was also
discussed in the presence of an electric field in superstring
theory ($Y=-1+E^2,$ $C^{11}=-1$) \cite{Sen:2003bc}.

When $C^{11}$ is positive ($k_1$ is pure imaginary),
the reality condition for the tachyon vertex operators in
Eq.~(\ref{tvo1}) requires that both $\lambda_{+}$ and $\lambda_{-}$
should be real. According to the boundary values of $T$,
the tachyon vertices are classified by three types:
(i) $T(-\infty)=\mp\infty$ and $T(+\infty)=\pm\infty$;
(ii) $T(-\infty)=0$ and $T(+\infty)=\pm\infty$;
(iii) $T(-\infty)=\pm\infty$ and $T(+\infty)=\pm\infty$:
\begin{equation}\label{v2}
T(X^1)=\left\{
\begin{array}{l}
\displaystyle{
\mbox{(i)}\,\,\,\,\lambda
\sinh \left( \kappa X^1\right) }\\
\displaystyle{
\mbox{(ii)}\,\,\,\lambda
\exp \left(\pm\kappa X^1\right) }\\
\displaystyle{
\mbox{(iii)}\,\,\lambda
\cosh \left( \kappa X^1\right) }
\end{array}
\right. ,
\end{equation}
where
\be\nn
i k_1 = \sqrt{\frac{-Y}{C^{11}}}\equiv\kappa.
\ee

Now we check the exact marginality of $T(X^1)$.
The operator product expansion of $T(X^1)$ with itself contains
only pole singularity at most,
\be
T(X^1(\tau_1)) T(X^1(\tau_2)) = \frac{\tilde K}{(\tau_1- \tau_2)^2}
+ (\mbox{regular}\,\, \mbox{terms}),
\nn \\
\ee
where $\tilde K= -\frac{1}{2}\lambda^2$, 0, and $\frac{1}{2}\lambda^2$
for (i), (ii), and (iii) in Eq.~(\ref{v2}), respectively.
(Note that in this calculation the noncommutative parameter $\theta^{\mu\nu}$
plays no role, since $T(X)$ consists only of a single field $X^1$.)
This implies that the operator $T(X)$ is self-local~\cite{Recknagel:1998ih},
which guarantees that it is exactly marginal.

\section{Tachyon Kinks with Electromagnetic field}
In the $\sigma$-model approach to the string theory, the
partition function of the worldsheet action with exact marginal
couplings is identified as the spacetime action, and the couplings are
interpreted as spacetime fields~\cite{fradkin-tseytlin}.
In relation to rolling
tachyons, the energy-momentum tensor has been obtained from the
partition function of the worldsheet action (\ref{SBCFT})
without gauge field interaction \cite{Larsen:2002wc}.
Here we take into account the gauge field interaction
on the worldsheet boundary induced by the constant electromagnetic
fields in addition to the tachyon vertex term.
The energy-momentum tensor in BCFT can be read from the
partition function of worldsheet theory coupled to
background gravity~\cite{Larsen:2002wc}:
\begin{equation}\label{staction}
S=Z_{{\rm disk}} \sim \int[dX^{\mu}]e^{-S_{\rm BCFT}},
\end{equation}
where $Z$ is the disk partition function, and
we have replaced $\eta_{\mu\nu}$ in $S_{\rm BCFT}$
with the generic curved spacetime metric $g_{\mu\nu}$.
From the spacetime action (\ref{staction})
we read the energy-momentum tensor in flat spacetime:
\begin{eqnarray}\nn
T_{\mu\nu}&\equiv& -\frac{2}{\sqrt{-g}}\left.
\frac{\delta S}{\delta g^{\mu\nu}}
\right|_{g_{\mu\nu}=\eta_{\mu\nu}}\\
&=&K \left[\eta_{\mu\nu}B(x) + A_{\mu\nu}(x)\right],
\label{EMT}
\end{eqnarray}
where
\begin{eqnarray}
B(x)&=& \int [dX^{'\mu}]e^{-[S_0(X^{'\mu})+S_{A}(X^{'\mu})
+S_{ T}(x^{\mu}+X^{'\mu})]}
\nn \\
&=& \left\langle e^{-S_{T} ( x^\mu+ X^{' \mu})}\right\rangle_{A} ,
\label{B}\\
A_{\mu\nu}(x)&=&\int [dX^{'\mu}]\,
\left(:\partial X^{'}_{\mu}(0)\bar\partial X^{'}_{\nu}(0):
\right.
\nn \\
&&\left.
\hskip 1.7cm +:\partial X^{'}_{\nu}(0)\bar\partial X^{'}_{\mu}(0):\right)
\nn \\
&&\hskip 1.2cm \times\,e^{-[S_{A}(X^{'\mu})+S_{T}(x^{\mu}+X^{'\mu})]}
\nn \\
&=& \left\langle \left(:\partial X^{'}_{\mu}(0)\bar\partial X^{'}_{\nu}(0):
+:\partial X^{'}_{\nu}(0)\bar\partial X^{'}_{\mu}(0):\right)\right.
\nn \\
&&\left. \times\, e^{-S_{T}(x^{\mu}+X^{'\mu})}\right\rangle_{A}.
\label{A}
\end{eqnarray}
The notations used here are as follows.
$K$ in Eq.~(\ref{EMT}) is an overall constant.
The symbol $:\,\,:$ denotes the normal ordering defined by
\bear\label{normord}
:\partial X^\mu (z) \bar\partial^{'} X^\nu (z'):
&=& \partial X^\mu (z) \bar\partial^{'} X^\nu (z')
\nn \\
&&+ \eta^{\mu\nu} \partial\bar\partial^{'}\log |z-z'|.
\eear
$X^{\mu}$ is split into the center of mass coordinate $x^{\mu}$ and
fluctuations $X^{'\mu}$, {\it i.e.,} $X^{\mu}=x^{\mu}+X^{'\mu}$.
Finally, $\langle \cdots \rangle_A$ denotes the vacuum expectation
value in the presence of the gauge field.
Proper normalization is made by choosing the vacuum-to-vacuum
expectation value for the unit operator in the presence of the
constant electromagnetic fields as
\be \label{vev}
\langle 1 \rangle_{A} = \sqrt{-\det (\eta_{\mu\nu} + F_{\mu\nu})}\, .
\ee

Now we assume the case where $p\ge2$ and $C^{11} > 0$. Then,
the operators in Eq.~(\ref{v2}) are exactly marginal and we can
consider the deformation by them. Here we will consider only the
exponential type,
\be\label{tvo3}
T(X^1) = \lambda \,e^{\kappa X^1},
\ee
and compute the spacetime energy-momentum tensor in Eq.~(\ref{EMT}).
After some calculations following the method in \cite{Larsen:2002wc}, we find
\bear\label{B3}
B(x^1) &=& \sqrt{-Y} \, f(x^1), \\
\label{A2}
A^{\mu\nu}(x^1)&=& 2 \sqrt{-Y}
\,\left[\left(G^{\mu\nu} - \frac12 \eta^{\mu\nu}
\right.\right.\nn \\
&&\hskip 1.2cm \left.\left. + \kappa^2\left(G^{\mu 1} G^{\nu 1}
- \theta^{\mu 1}\theta^{\nu 1}\right)
\right.\Big) f(x^1) \right.
\nn \\
&& \hskip 1.1cm \left.- \kappa^2 \left(G^{\mu 1} G^{\nu 1}
- \theta^{\mu 1}\theta^{\nu 1}\right)\right.\Big],
\eear
where
\be\label{f0}
f(x^1) \equiv \frac1{1+2\pi\lambda e^{\kappa x^1}}.
\ee

Combining Eqs.~(\ref{B3}), (\ref{A2}), and (\ref{EMT}), we finally obtain
\bear\label{EMT2} T^{\mu\nu}
&=& {\cal T}_{p}\sqrt{-Y} \kappa^2\left[-\left(\frac1{\kappa^2}
G^{\mu\nu}+ G^{\mu 1} G^{\nu 1} - \theta^{\mu 1}\theta^{\nu
1}\right) f
\right.\nn \\
&&\left.\hskip 2cm +G^{\mu 1} G^{\nu 1} -\theta^{\mu 1}\theta^{\nu 1}\right. \Big],
\eear
where we have determined the overall normalization constant,
$K=-\frac12 {\cal T}_{p}$, by comparison with the static limit,
\be
T^{00}(\lambda=0, F_{\mu\nu}=0) = {\cal T}_{p}.
\ee
Note that the energy-momentum tensor (\ref{EMT2}) satisfies the
conservation law,
\be\label{EMT3}
\partial_{\mu} T^{\mu\nu} = 0.
\ee
In addition, we compute the source for the antisymmetric tensor field
from the spacetime action (\ref{staction}) as
\bear\label{FD}
\Pi^{\mu\nu} &\equiv& 2 \frac{\delta S}{\delta F_{\mu\nu}}
\nn \\
&=& \frac{{\cal T}_{p}}{2}
\left\langle \left(:\partial X^{'}_{\mu}(0)\bar\partial X^{'}_{\nu}(0):
 -:\partial X^{'}_{\nu}(0)\bar\partial X^{'}_{\mu}(0):\right)\right.
\nn \\
&& \left.\hskip 0.5cm  \times\, e^{-\lambda\int d\tau e^{\kappa (x^1+X^{'1}) }}
\right\rangle_{A}
\nn \\
&=& {\cal T}_{p} \sqrt{-Y}\kappa^2\left\{\left[\frac1{\kappa^2}\,
\theta^{\mu\nu} - \left(G^{\mu 1}\theta^{\nu 1} - G^{\nu 1}\theta^{\mu 1}
\right)\right]\, f \right.
\nn \\
&& \hskip 1.7cm \left.+ G^{\mu 1}\theta^{\nu 1}
- G^{\nu 1}\theta^{\mu 1}\right.\Big\}.
\eear
{}From Eqs.~(\ref{EMT2}) and (\ref{FD}), we see that both
$T^{\mu\nu}$ and $\Pi^{\mu\nu}$ have essentially the same
$x^1$-dependent term given by Eq.~(\ref{f0}). For some components,
the $x^1$-dependent part vanishes. Let us consider the case $p=2$
for definiteness.
Then, $-Y=1-E_{1}^{2}-E_{2}^{2}+B^{2} (>0)$ and $C^{11} = -1+E_2^2 (>0)$
and, we find constant components
\begin{eqnarray}
&&\frac{\Pi^{01}}{E_{1}}=\frac{\Pi^{12}}{B}=\frac{T^{01}}{E_{2}B}
=\frac{T^{11}}{1-E_{2}^{2}}=-\frac{T^{02}}{E_{1}B}=-\frac{T^{12}}{E_{1}E_{2}}
\nonumber\\
&&=\frac{{\cal T}_{2}}{\sqrt{-Y}},
\label{c}
\end{eqnarray}
while the other components are given by the sums of an $x^1$-dependent piece
and a constant,
\begin{eqnarray}
\Pi^{02}
&=&-\frac{{\cal T}_{2}\sqrt{-Y}E_{2}}{1-E_{2}^{2}}f
    +\frac{E_{2}(E_{1}^{2}-B^{2})}{E_{1}(1-E_{2}^{2})}\Pi^{01} ,
\label{pi2}\\
T^{00}
&=&\frac{{\cal T}_{2}\sqrt{-Y}}{1-E_{2}^{2}}f
+\frac{E_{1}^{2}+E_{2}^{2}B^{2}}{E_{1}(1-E_{2}^{2})}\Pi^{01} ,
\label{t00}\\
T^{22}
&=& \frac{{\cal T}_{2}\sqrt{-Y}}{1-E_{1}^{2}}f
-\frac{E_{1}^{2}E_{2}^{2}-B^{2}}{E_{1}(1-E_{1}^{2})}\Pi^{01} .
\label{t22}
\end{eqnarray}

Note that for positive $\lambda$, $f(x^1)$ is regular everywhere, while
it has a singular point if $\lambda$ is negative. This is a well-known
feature of bosonic theory and implies an instability of bosonic
theory~\cite{Sen:2002nu} in which the effective tachyon potential
is unbounded from below in the negative tachyon direction.
For positive $\lambda$, the static tachyon configuration connects
the perturbative string vacuum at $T=0$ and the vacuum at $T=\infty$, where
the unstable D-brane disappears. Therefore, it may be interpreted as
a {\it half brane} ($\frac{1}{2}$D$(p-1)$-brane) similar to a bubble wall
at a given time connecting two phases.
This configuration has already been obtained
as a half kink in DBI-type EFT~\cite{Kim:2003in}, NCFT~\cite{Banerjee:2004cw},
and BSFT~\cite{Kim:2006mg}. Since these effective actions can be derived
from the superstring theory, we expect that the half brane can also be
obtained in super BCFT.

In addition to the exponential-type vertex, there
exist two more boundary tachyon operators, hyperbolic sine and cosine in
Eq.~(\ref{v2}). They could also be studied in the context of BCFT.

\vspace{5mm}

\section{Concluding Remarks}

We have investigated tachyon vertex operators
which are exactly marginal and depend on only one spatial direction $X^1$
in the presence of background electromagnetic fields
in bosonic string theory.

We found that, when $\det (\eta + F)<0$ and $C^{11}>0$, there are
hyperbolic sine, hyperbolic cosine, and exponential-type
 exactly marginal operators. This case can be realized
depending on the magnitude of electric and magnetic
fields along the transverse direction of $x^1$.
Therefore, the dimension of the unstable D-brane should be equal to
or larger than two.

We obtained the general form of the energy-momentum
tensor for the exponential-type operator by using
the path integral approach developed in Ref.~\cite{Larsen:2002wc}.
It may be interpreted as a half brane, which has been found in
other approaches such as DBI-type EFT, NCFT, and BSFT.

Our analysis may be extended to the superstring case and also to other marginal
operators, {\it i.e.,} the hyperbolic sine and hyperbolic cosine cases.
It may also be intriguing to include the R-R coupling~\cite{Afonso:2006ws}.

\begin{acknowledgments}
We would like to thank Y. Okawa, M. Schnabl, and A. Sen for helpful
discussions.
This work was supported by the Korea Research Foundation Grant
funded by the Korean Government (MOEHRD) (KRF-2005-005-J11903) (A.I.)
and by the Science Research Center Program of
the Korea Science and Engineering Foundation through the Center for
Quantum Spacetime (CQUeST) of Sogang University with Grant No.
R11-2005-021 (C.K. and O.K.). This work is also the result of research
activities (Astrophysical Research Center for the Structure and
Evolution of the Cosmos (ARCSEC)) and Grant No. R01-2006-000-10965-0
from the Basic Research Program supported by the Korea Science $\&$
Engineering Foundation (Y.K.).
\end{acknowledgments}



\end{document}